\documentclass[12pt]{article}
\usepackage{amsmath,amsthm}
\usepackage{eucal}
\usepackage{eufrak}
\usepackage{pb-diagram,lamsarrow,pb-lams}

\DeclareMathAlphabet{\bb}{U}{msb}{m}{n} \gdef\C{\bb C} \gdef\dZ{\bb
Z}    \gdef\R{\bb R}
\gdef\K{\bb K} \gdef\BH{\bb H} \gdef\F{\bb F} 

 \DeclareMathOperator{\spin}{{\bf
Spin}} \DeclareMathOperator{\pin}{{\bf Pin}}

\DeclareMathOperator{\Id}{Id} 
  
\DeclareMathOperator{\Ker}{Ker}

\DeclareMathOperator{\Ext}{Ext}

\newcommand{\cA}{\mathcal{A}}

\newcommand{\sA}{{\sf A}}

\newcommand{\sI}{{\sf I}}
\newcommand{\sW}{{\sf W}}
\newcommand{\sE}{{\sf E}}
\newcommand{\sC}{{\sf C}}
\newcommand{\sF}{{\sf F}}
\newcommand{\sT}{{\sf T}}
\newcommand{\sS}{{\sf S}}

\newcommand{\sK}{{\sf K}}

\newcommand{\cl}{C\kern -0.2em \ell}

\begin{document}
\title{$CPT$ groups for spinor field in de Sitter space}
\author{V. V. Varlamov\\
{\small\it Department of Mathematics, Siberia State University of Industry,}\\
{\small\it Kirova 42, Novokuznetsk 654007, Russia}}
\date{}
\maketitle
\begin{abstract}
A group structure of the discrete transformations (parity, time
reversal and charge conjugation) for spinor field in de Sitter space
are studied in terms of extraspecial finite groups. Two $CPT$ groups
are introduced, the first group from an analysis of the de
Sitter-Dirac wave equation for spinor field, and the second group
from a purely algebraic approach based on the automorphism set of
Clifford algebras. It is shown that both groups are isomorphic to
each other.
\end{abstract}
\section{Introduction}
Quantum field theory in de Sitter spacetime has been extensively
studied during the past two decades with the purpose of
understanding the generation of cosmic structure from inflation and
the problems surrounding the cosmological constant. It is well known
that discrete symmetries play an important role in the standard
quantum field theory in Minkowski spacetime. In recent paper
\cite{MRT05} discrete symmetries for spinor field in de Sitter space
with the signature $(+,-,-,-,-)$ have been derived from the analysis
of the de Sitter-Dirac wave equation. Discrete symmetries in de
Sitter space with the signature $(+,+,+,+,-)$ have been considered
in the work \cite{Var04} within an algebraic approach based on the
automorphism set of Clifford algebras.

In the present paper we study a group structure of the discrete
transformations in the framework of extraspecial finite groups. In
the section 3 we introduce a $CPT$ group for discrete symmetries in
the representation of the work \cite{MRT05} (for more details about
$CPT$ groups see \cite{Var01,Var04a,Var04b,Soc04}). It is shown that
the discrete transformations form a non-Abelian finite group of
order 16. Group isomorphisms and order structure are elucidated for
this group. Other realization of the $CPT$ group is given in the
section 4. In this section we consider an automorphism set  of the
Clifford algebra associated with the de Sitter space. It is proven
that a $CPT$ group, formed within the automorphism set, is
isomorphic the analogous group considered in the section 3.
\section{Preliminaries}
Usually, the de Sitter space is understood as a hyperboloid embedded
in a five-dimensional Minkowski space $\R^{1,4}$
\begin{equation}\label{dS}
X_H=\left\{x\in\R^{1,4}:\;x^2=\eta_{\alpha\beta}x^\alpha
x^\beta=-H^{-2}\right\},\quad \alpha,\beta=0,1,2,3,4,
\end{equation}
where $\eta_{\alpha\beta}=\text{diag}(1,-1,-1,-1,-1)$.

The spinor wave equation in the de Sitter
space-time\footnote{Originally, relativistic wave equations in a
five-dimensional pseudoeuclidean space (de Sitter space) were
introduced by Dirac in 1935 \cite{Dir35}. They have the form
$(i\gamma_\mu\partial_\mu+m)\psi=0$, where five $4\times 4$ Dirac
matrices form the Clifford algebra $\cl_{4,1}$.} (\ref{dS}) has been
given in the works \cite{BGMT01,MRT05}, and can be derived via the
eigenvalue equation for the second order Casimir operator,
\begin{equation}\label{dSDir}
(-i\not\! x\gamma\cdot\overline{\partial}+2i+\nu)\psi(x)=0,
\end{equation}
where $\not\! x=\eta^{\alpha\beta}\gamma_\alpha x_\beta$ and
$\overline{\partial}_\alpha=\partial_\alpha+H^2x_\alpha
x\cdot\partial$. In this case the $4\times 4$ matrices
$\gamma_\alpha$ are spinor representations of the units of the
Clifford algebra $\cl_{1,4}$ and satisfy the relations
\[
\gamma_\alpha\gamma_\beta+\gamma_\beta\gamma_\alpha=2\eta_{\alpha\beta}
\boldsymbol{1}.
\]
An explicit representation\footnote{In general, the Clifford algebra
$\cl_{1,4}$, associated with the de Sitter space $\R^{1,4}$, has a
double quaternionic ring $\K=\BH\oplus\BH$ \cite{Lou91}, the type
$p-q\equiv 5\pmod{8}$. For this reason, the algebra $\cl_{1,4}$
admits the following decomposition into a direct sum:
$\cl_{1,4}\simeq\cl_{1,3}\oplus\cl_{1,3}$, where $\cl_{1,3}$ is a
{\it spacetime algebra}. There is a homomorphic mapping
$\epsilon:\,\cl_{1,4}\rightarrow{}^\epsilon\cl_{1,3}$, where
${}^\epsilon\cl_{1,3}\simeq\cl_{1,3}/\Ker\epsilon$ is a quotient
algebra, $\Ker\epsilon$ is a kernel of $\epsilon$. The basis
(\ref{Gamma}) is one from the set of isomorphic spinbasises obtained
via the homomorphism $\epsilon$.} for $\gamma_\alpha$ chosen in
\cite{MRT05} is
\begin{gather}
\gamma_0=\begin{pmatrix} \boldsymbol{1}_2 & 0\\
0 & -\boldsymbol{1}_2
\end{pmatrix},\quad
\gamma_1=\begin{pmatrix} 0 & i\sigma_1\\
i\sigma_1 & 0
\end{pmatrix},\nonumber\\
\gamma_2=\begin{pmatrix} 0 & -i\sigma_2\\
-i\sigma_2 & 0
\end{pmatrix},\quad
\gamma_3=\begin{pmatrix} 0 & i\sigma_3\\
i\sigma_3 & 0
\end{pmatrix},\nonumber\\
\gamma_4=\begin{pmatrix} 0 & \boldsymbol{1}_2\\
-\boldsymbol{1}_2 & 0
\end{pmatrix},\label{Gamma}
\end{gather}
where $\boldsymbol{1}_2$ is a $2\times 2$ unit and $\sigma_i$ are
Pauli matrices.

Discrete symmetries (parity transformation $P$, time reversal $T$
and charge conjugation $C$), obtained from analysis of the equation
(\ref{dSDir}), in spinor notation have the form \cite{MRT05}
\begin{equation}\label{DT}
P=\eta_p\gamma_0\gamma_4,\quad T=\eta_t\gamma_0,\quad
C=\eta_c\gamma_2,
\end{equation}
where $\eta_p,\,\eta_t,\,\eta_c$ are arbitrary unobservable phase
quantities.
\section{The $CPT$ group}
In this section we will show that the transformations (\ref{DT})
form a finite group of order 16, a so called $CPT$ group. Moreover,
this group is a subgroup of the more large finite group associated
with the algebra $\cl_{1,4}$.

As is known \cite{Sal81a,Sal82,Sal84,Bra85}, a structure of the
Clifford algebras admits a very elegant description in terms of
finite groups. In accordance with a multiplication rule
\begin{equation}\label{e1}
\gamma^2_i=\sigma(p-i)\boldsymbol{1},\quad\gamma_i\gamma_j=-\gamma_j\gamma_i,
\end{equation}
\begin{equation}\label{e2}
\sigma(n)=\left\{\begin{array}{rl}
-1 & \mbox{if $n\leq 0$},\\
+1 & \mbox{if $n>0$},
\end{array}\right.
\end{equation}
basis elements  of the Clifford algebra $\cl_{p,q}$ (the algebra
over the field of real numbers, $\F=\R$) form a finite group of
order $2^{n+1}$,
\begin{equation}\label{FG}
G(p,q)=\left\{\pm
1,\,\pm\gamma_i,\,\pm\gamma_i\gamma_j,\,\pm\gamma_i\gamma_j\gamma_k,\,\ldots,\,
\pm\gamma_1\gamma_2\cdots\gamma_n\right\}\quad(i<j<k<\ldots).
\end{equation}
The finite group $G(1,4)$, associated with the algebra $\cl_{1,4}$,
is a particular case of (\ref{FG}),
\[
G(1,4)=\left\{\pm 1,\,\pm\gamma_0,\,\pm\gamma_1,\,\ldots,\,
\pm\gamma_0\gamma_1\gamma_2\gamma_3\gamma_4\right\},
\]
where $\gamma_i$ have the form (\ref{Gamma}). It is a finite group
of order 64 with an order structure (23,40). Moreover, $G(1,4)$ is
an {\it extraspecial two-group} \cite{Sal84,Bra85}. In Salingaros
notation the following isomorphism holds:
\[
G(1,4)=\Omega_4\simeq N_4\circ(\dZ_2\otimes\dZ_2)\simeq Q_4\circ
D_4\circ(\dZ_2\otimes\dZ_2),
\]
where $Q_4$ is a quaternion group, $D_4$ is a dihedral group,
$\dZ_2\otimes\dZ_2$ is a Gauss-Klein viergroup, $\circ$ is a {\it
central product} ($Q_4$ and $D_4$ are finite groups of order 8).

As is known, the orthogonal group $O(p,q)$ of the real space
$\R^{p,q}$ is represented by a semidirect product
$O_0(p,q)\odot\{1,\,P,\,T,\,PT\}$, where $O_0(p,q)$ is a connected
component, $\{1,\,P,\,T,\,PT\}$ is a discrete subgroup (reflection
group). If we take into account the charge conjugation $C$, then we
come to the product
$O_0(p,q)\odot\{1,\,P,\,T,\,PT,\,C,\,CP,\,CT,\,CPT\}$. Universal
coverings of the groups $O(p,q)$ are Clifford-Lipschitz groups
$\pin(p,q)$ which are completely constructed within the Clifford
algebras $\cl_{p,q}$ \cite{Lou91}. It has been recently shown
\cite{Var04a,Var04b} that there exist 64 universal coverings of the
orthogonal group $O(p,q)$:
\[
\pin^{a,b,c,d,e,f,g}(p,q)\simeq\frac{(\spin_+(p,q)\odot
C^{a,b,c,d,e,f,g})}{\dZ_2},
\]
where
\[
C^{a,b,c,d,e,f,g}=\{\pm 1,\,\pm P,\,\pm T,\,\pm PT,\,\pm C,\,\pm
CP,\, \pm CT,\,\pm CPT\}
\]
is {\it a full $CPT$-group}. $C^{a,b,c,d,e,f,g}$ is a finite group
of order 16 (a complete classification of these groups is given in
\cite{Var04b}). At this point, the group
\[
\Ext(\cl_{p,q})=\frac{C^{a,b,c,d,e,f,g}}{\dZ_2}
\]
is called {\it a generating group}.

Let us define a $CPT$ group for the spinor field in de Sitter space.
The invariance of the dS-Dirac equation (\ref{dSDir}) with respect
to $P$-, $T$-, and $C$- transformations leads to the representation
(\ref{DT}). For simplicity we suppose that all the phase quantities
are equal to the unit, $\eta_p=\eta_t=\eta_c=1$. Thus, we can form a
finite group of order 8
\begin{equation}
\{1,\,P,\,T,\,PT,\,C,\,CP,\,CT,\,CPT\}
\sim\{1,\,\gamma_0\gamma_4,\,\gamma_0,\,\gamma_4,\,\gamma_2,\,
\gamma_0\gamma_2\gamma_4,\,\gamma_0\gamma_2,\,\gamma_2\gamma_4\}.
\label{CPT1}
\end{equation}
It is easy to verify that a multiplication table of this group has
the form
\begin{center}{\renewcommand{\arraystretch}{1.4}
\begin{tabular}{|c||c|c|c|c|c|c|c|c|}\hline
  & $1$ & $\gamma_{04}$ & $\gamma_{0}$ & $\gamma_{4}$ & $\gamma_{2}$ &
$\gamma_{024}$ & $\gamma_{02}$ & $\gamma_{24}$\\ \hline\hline $1$ &
$1$ & $\gamma_{04}$ & $\gamma_{0}$ & $\gamma_{4}$ & $\gamma_2$ &
$\gamma_{024}$ & $\gamma_{02}$ & $\gamma_{24}$\\ \hline
$\gamma_{04}$ & $\gamma_{04}$ & $1$ & $-\gamma_{4}$ & $-\gamma_{0}$
&
$-\gamma_{024}$ & $-\gamma_{2}$ & $\gamma_{24}$ & $\gamma_{02}$\\
\hline $\gamma_{0}$ & $\gamma_{0}$ & $\gamma_{4}$ & $1$ &
$\gamma_{04}$
& $\gamma_{02}$ & $\gamma_{24}$ & $\gamma_{2}$ & $\gamma_{024}$\\
\hline $\gamma_{4}$ & $\gamma_{4}$ & $\gamma_{0}$ & $-\gamma_{04}$ &
 $-1$ & $-\gamma_{24}$ & $-\gamma_{02}$ & $\gamma_{024}$ &
$\gamma_{2}$\\ \hline $\gamma_{2}$ & $\gamma_{2}$ & $-\gamma_{024}$
& $-\gamma_{02}$ & $\gamma_{24}$ & $-1$ & $\gamma_{04}$ &
$\gamma_{0}$ & $-\gamma_{4}$\\ \hline $\gamma_{024}$ &
$\gamma_{024}$ & $-\gamma_{2}$ & $\gamma_{24}$ & $-\gamma_{02}$ &
$\gamma_{04}$ & $-1$ & $\gamma_{4}$ & $-\gamma_{0}$\\ \hline
$\gamma_{02}$ & $\gamma_{02}$ & $-\gamma_{24}$ & $-\gamma_{2}$
& $\gamma_{024}$ & $-\gamma_{0}$ & $\gamma_{4}$ & $1$ & $-\gamma_{04}$\\
\hline $\gamma_{24}$ & $\gamma_{24}$ & $-\gamma_{02}$ &
$\gamma_{024}$ & $-\gamma_{2}$ & $\gamma_{4}$ & $-\gamma_{0}$ &
$\gamma_{04}$ & $-1$\\ \hline
\end{tabular}.
}
\end{center}
Here $\gamma_{04}\equiv\gamma_0\gamma_4$,
$\gamma_{024}\equiv\gamma_0\gamma_2\gamma_4$ and so on. Hence it
follows that the group (\ref{CPT1}) is a non-Abelian finite group of
the order structure (3,4). In more details, it is the group
$\overset{\ast}{\dZ}_4\otimes\dZ_2$ with the signature
$(+,+,-,-,-,+,-)$. Therefore, the $CPT$ group in de Sitter spacetime
is
\[
C^{+,+,-,-,-,+,-}\simeq\overset{\ast}{\dZ}_4\otimes\dZ_2\otimes\dZ_2.
\]
It is easy to see that $C^{+,+,-,-,-,+,-}$ is a subgroup of
$G(1,4)$. In this case, the universal covering of the de Sitter
group is defined as
\[
\pin^{+,+,-,-,-,+,-}(1,4)\simeq
\frac{(\spin_+(1,4)\odot\overset{\ast}{\dZ}_4\otimes\dZ_2\otimes\dZ_2)}{\dZ_2}.
\]
\section{Discrete symmetries and automorphisms of the Clifford
algebras} Within the Clifford algebra $\C_n$ (the algebra over the
field of complex numbers, $\F=\C$) there exist eight automorphisms
\cite{Ras55,Var04a} (including an identical automorphism $\Id$). We
list these transformations and their spinor representations:
\begin{eqnarray}
\cA\longrightarrow\cA^\star,&&\quad\sA^\star=\sW\sA\sW^{-1},\nonumber\\
\cA\longrightarrow\widetilde{\cA},&&\quad\widetilde{\cA}=\sE\sA^{\sT}\sE^{-1},
\nonumber\\
\cA\longrightarrow\widetilde{\cA^\star},&&\quad\widetilde{\sA^\star}=
\sC\sA^{\sT}\sC^{-1},\quad\sC=\sE\sW,\nonumber\\
\cA\longrightarrow\overline{\cA},&&\quad\overline{\sA}=\Pi\sA^\ast\Pi^{-1},
\nonumber\\
\cA\longrightarrow\overline{\cA^\star},&&\quad\overline{\sA^\star}=
\sK\sA^\ast\sK^{-1},\quad\sK=\Pi\sW,\nonumber\\
\cA\longrightarrow\overline{\widetilde{\cA}},&&\quad
\overline{\widetilde{\sA}}=\sS\left(\sA^{\sT}\right)^\ast\sS^{-1},\quad
\sS=\Pi\sE,\nonumber\\
\cA\longrightarrow\overline{\widetilde{\cA^\star}},&&\quad
\overline{\widetilde{\sA^\star}}=\sF\left(\sA^\ast\right)^{\sT}\sF^{-1},\quad
\sF=\Pi\sC,\nonumber
\end{eqnarray}
where the symbol $\sT$ means a transposition, and $\ast$ is a
complex conjugation. In general, the real algebras $\cl_{p,q}$ also
admit all the eight automorphisms, excluding the case $p-q\equiv
0,1,2\pmod{8}$ when a pseudoautomorphism
$\cA\rightarrow\overline{\cA}$ is reduced to the identical
automorphism $\Id$. It is easy to verify that an automorphism set
$\Ext(\C_n)=\{\Id,\,\star,\,\widetilde{\phantom{cc}},\,\widetilde{\star},\,
\overline{\phantom{cc}},\,\overline{\star},\,
\overline{\widetilde{\phantom{cc}}},\,\overline{\widetilde{\star}}\}$
of $\C_n$ forms a finite group of order 8. Moreover, there is an
isomorphism between $\Ext(\C_n)$ and a $CPT$--group of the discrete
transformations,
$\Ext(\C_n)\simeq\{1,\,P,\,T,\,PT,\,C,\,CP,\,CT,\,CPT\}$. In this
case, space inversion $P$, time reversal $T$, full reflection $PT$,
charge conjugation $C$, transformations $CP$, $CT$ and the full
$CPT$--transformation correspond to the automorphism
$\cA\rightarrow\cA^\star$, antiautomorphisms
$\cA\rightarrow\widetilde{\cA}$,
$\cA\rightarrow\widetilde{\cA^\star}$, pseudoautomorphisms
$\cA\rightarrow\overline{\cA}$,
$\cA\rightarrow\overline{\cA^\star}$, pseudoantiautomorphisms
$\cA\rightarrow\overline{\widetilde{\cA}}$ and
$\cA\rightarrow\overline{\widetilde{\cA^\star}}$, respectively (for
more details, see \cite{Var04a}).

Let us study an automorphism group of the algebra $\cl_{1,4}$. First
of all, $\cl_{1,4}$ has the type $p-q\equiv 5\pmod{8}$, therefore,
all the eight automorphisms exist. Using the $\gamma$-matrices of
the basis (\ref{Gamma}), we will define elements of the group
$\Ext(\cl_{1,4})$. At first, the matrix of the automorphism
$\cA\rightarrow\cA^\star$ has the form
$\sW=\gamma_0\gamma_1\gamma_2\gamma_3\gamma_4\equiv\gamma_{01234}$.
Further, since
\begin{gather}
\gamma^{\sT}_0=\gamma_0,\quad\gamma^{\sT}_1=\gamma_1,\quad
\gamma^{\sT}_2=-\gamma_2,\nonumber\\
\gamma^{\sT}_3=\gamma_3,\quad\gamma^{\sT}_4=-\gamma_4,\nonumber
\end{gather}
then in accordance with $\widetilde{\sA}=\sE\sA^{\sT}\sE^{-1}$ we
have
\begin{gather}
\gamma_0=\sE\gamma_0\sE^{-1},\quad\gamma_1=\sE\gamma_1\sE^{-1},\quad
\gamma_2=-\sE\gamma_2\sE^{-1},\nonumber\\
\gamma_3=\sE\gamma_3\sE^{-1},\quad\gamma_4=-\sE\gamma_4\sE^{-1}.
\nonumber
\end{gather}
Hence it follows that $\sE$ commutes with $\gamma_0$, $\gamma_1$,
$\gamma_3$ and anticommutes with $\gamma_2$ and $\gamma_4$, that is,
$\sE=\gamma_2\gamma_4$. From the definition $\sC=\sE\sW$ we find
that a matrix of the antiautomorphism
$\cA\rightarrow\widetilde{\cA^\star}$ has the form
$\sC=\gamma_0\gamma_1\gamma_3$. The basis (\ref{Gamma}) contains
both complex and real matrices:
\begin{gather}
\gamma^\ast_0=\gamma_0,\quad\gamma^\ast_1=-\gamma_1,\quad
\gamma^\ast_2=\gamma_2,\nonumber\\
\gamma^\ast_3=-\gamma_3,\quad\gamma^\ast_4=\gamma_4.\nonumber
\end{gather}
Therefore, from $\overline{\sA}=\Pi\sA^\ast\Pi^{-1}$ we have
\begin{gather}
\gamma_0=\Pi\gamma_0\Pi^{-1},\quad\gamma_1=-\Pi\gamma_1\Pi^{-1},\quad
\gamma_2=\Pi\gamma_2\Pi^{-1},\nonumber\\
\gamma_3=-\Pi\gamma_3\Pi^{-1},\quad\gamma_4=\Pi\gamma_4\Pi^{-1}.\nonumber
\end{gather}
From the latter relations we obtain $\Pi=\gamma_1\gamma_3$. Further,
in accordance with $\sK=\Pi\sW$ for the matrix of the
pseudoautomorphism $\cA\rightarrow\overline{\cA^\star}$ we have
$\sK=\gamma_0\gamma_2\gamma_4$. Finally, for the
pseudoantiautomorphisms $\cA\rightarrow\overline{\widetilde{\cA}}$,
$\cA\rightarrow\overline{\widetilde{\cA^\star}}$ from the
definitions $\sS=\Pi\sE$, $\sF=\Pi\sC$ it follows that
$\sS=\gamma_1\gamma_2\gamma_3\gamma_4$, $\sF=\gamma_0$. Thus, we
come to the following automorphism group:
\begin{multline}
\Ext(\cl_{1,4})\simeq\{\sI,\,\sW,\,\sE,\,\sC,\,\Pi,\,\sK,\,\sS,\,\sF\}\simeq\\
\simeq\{1,\,\gamma_0\gamma_1\gamma_2\gamma_3\gamma_4,\,\gamma_2\gamma_4,\,
\gamma_0\gamma_1\gamma_3,\,\gamma_1\gamma_3,\,\gamma_0\gamma_2\gamma_4,\,
\gamma_1\gamma_2\gamma_3\gamma_4,\,\gamma_0\}. \label{CPT2}
\end{multline}
The multiplication table of this group has the form
\begin{center}{\renewcommand{\arraystretch}{1.4}
\begin{tabular}{|c||c|c|c|c|c|c|c|c|}\hline
  & $1$ & $\gamma_{01234}$ & $\gamma_{24}$ & $\gamma_{013}$ & $\gamma_{13}$ &
$\gamma_{024}$ & $\gamma_{1234}$ & $\gamma_{0}$\\ \hline\hline $1$ &
$1$ & $\gamma_{01234}$ & $\gamma_{24}$ & $\gamma_{013}$ &
$\gamma_{13}$ & $\gamma_{024}$ & $\gamma_{1234}$ & $\gamma_{0}$\\
\hline $\gamma_{01234}$ & $\gamma_{01234}$ & $1$ & $\gamma_{013}$ &
$\gamma_{24}$ &
$\gamma_{024}$ & $\gamma_{13}$ & $\gamma_{0}$ & $\gamma_{1234}$\\
\hline $\gamma_{24}$ & $\gamma_{24}$ & $-\gamma_{013}$ & $-1$ &
$-\gamma_{01234}$
& $-\gamma_{1234}$ & $-\gamma_{0}$ & $\gamma_{13}$ & $\gamma_{024}$\\
\hline $\gamma_{013}$ & $\gamma_{013}$ & $\gamma_{24}$ &
$-\gamma_{01234}$ &
 $-1$ & $-\gamma_{0}$ & $-\gamma_{1234}$ & $\gamma_{024}$ &
$\gamma_{13}$\\ \hline $\gamma_{13}$ & $\gamma_{13}$ &
$\gamma_{024}$ & $-\gamma_{1234}$ & $-\gamma_{0}$ & $-1$ &
$-\gamma_{01234}$ & $\gamma_{24}$ & $\gamma_{013}$\\ \hline
$\gamma_{024}$ & $\gamma_{024}$ & $\gamma_{13}$ & $-\gamma_{0}$ &
$-\gamma_{1234}$ & $-\gamma_{01234}$ & $-1$ & $-\gamma_{013}$ & $\gamma_{24}$\\
\hline $\gamma_{1234}$ & $\gamma_{1234}$ & $\gamma_{0}$ &
$\gamma_{13}$
& $\gamma_{024}$ & $\gamma_{24}$ & $\gamma_{013}$ & $1$ & $\gamma_{01234}$\\
\hline $\gamma_{0}$ & $\gamma_{0}$ & $\gamma_{1234}$ &
$\gamma_{024}$ & $\gamma_{13}$ & $\gamma_{013}$ & $\gamma_{24}$ &
$\gamma_{01234}$ & $1$\\ \hline
\end{tabular}.
}
\end{center}
As follows from this table, the group $\Ext(\cl_{1,4})$ is
non-Abelian. More precisely, the group (\ref{CPT2}) is a finite
group $\overset{\ast}{\dZ}_4\otimes\dZ_2$ with the signature
$(+,-,-,-,-,+,+)$. In this case we have the following universal
covering:
\[
\pin^{+,-,-,-,-,+,+}(1,4)\simeq\frac{(\spin_+(1,4)\odot
C^{+,-,-,-,-,+,+})}{\dZ_2},
\]
where
\[
C^{+,-,-,-,-,+,+}\simeq\overset{\ast}{\dZ}_4\otimes\dZ_2\otimes\dZ_2
\]
is a full $CPT$ group of the spinor field in de Sitter space
$\R^{1,4}$. In turn, $C^{+,-,-,-,-,+,+}$ is a subgroup of $G(1,4)$.

Moreover, we see that the generating group (\ref{CPT1}) and
(\ref{CPT2}) are isomorphic,
\[
\{1,\,P,\,T,\,PT,\,C,\,CP,\,CT,\,CPT\}\simeq
\{\sI,\,\sW,\,\sE,\,\sC,\,\Pi,\,\sK,\,\sS,\,\sF\}\simeq
\overset{\ast}{\dZ}_4\otimes\dZ_2.
\]
Thus, we come to the following result: the finite group
(\ref{CPT1}), derived from the analysis of invariance properties of
the dS-Dirac equation with respect to discrete transformations $C$,
$P$ and $T$, is isomorphic to the automorphism group of the algebra
$\cl_{1,4}$. This result allows us to study discrete symmetries and
their group structure for physical fields without handling to
analysis of relativistic wave equations.

\end{document}